\documentclass[12pt]{article}
\usepackage[symbol]{footmisc}
\def\correspondingauthor{\footnote{Corresponding author.  }}
\usepackage[left=2cm,top=2.5cm,right=2cm,bottom=2.5cm]{geometry}
\usepackage{amsmath, amssymb}

\usepackage{graphicx}
\usepackage{graphics}
\usepackage{epstopdf}
\usepackage{subfigure}
\usepackage{amsfonts}
\usepackage{sectsty}
\usepackage{sectsty}
\usepackage{hyperref}
\usepackage{cite, multirow,hhline}

\begin{document}
	\begin{center}
	\large{\bf{Validation of Energy Conditions in Wormhole Geometry within Viable $f(R)$ Gravity}} \\
	\vspace{5mm}
	\normalsize{Gauranga C. Samanta$^1$ and Nisha Godani$^{2,}{}$\correspondingauthor{}, }\\
	\normalsize{$^1$ Department of Mathematics, BITS Pilani K K Birla Goa Campus, Goa, India\\
	$^2$ Department of Mathematics, Institute of Applied Sciences and Humanities\\ GLA University, Mathura, Uttar Pradesh, India}\\
	\normalsize {gauranga81@gmail.com \\nishagodani.dei@gmail.com }
\end{center}
	\begin{abstract}
		In this work, wormholes, tunnel like structures introduced by Morris \& Thorne \cite{Morris95}, are explored within the framework of $f(R)$ gravity. Using the shape function $b(r)=r_0\big(\frac{r}{r_0}\big)^\gamma$, where $0<\gamma<1$, and the equation of state $p_r=\omega\rho$, the $f(R)$ function is derived and the field equations are solved. Then null, weak, strong and dominated energy conditions are analyzed and spherical regions satisfying these energy conditions are determined. Furthermore, we calculated the range of the radius of the throat of the wormhole, where the energy conditions are satisfied.
	\end{abstract}

\section{Introduction}
The idea of traversable wormholes was initiated by Morris and Thorne\cite{Morris95}, they used the principle of general relativity to study the possibility of time travel for humans.
The general theory of relativity envisages that the geometry and structure of space-time in the presence of matter is not unbending but is flexible and deformable. The denser the body is, the stronger the curvature of space is, which basically indicates to the notion of black holes.
Nevertheless, in the latter case, the stuff of space-time drops its meaning at the curvature singularity. If somehow the creation of the singularity is avoided, then it would be possible to travel through the throat, so that there is no constraint to observer's motion on the manifold.
After, finding of the general theory of relativity, the possibility of such solution to the Einstein field equations was first time investigated by Flamm\cite{Flamm48}. Subsequently, it was shown that his solution was unstable. However, the more detailed solutions of the wormhole was studied by Einstein and Rosen\cite{Einstein73}.
A wormhole is a topological feature of the space-time, which has basically been considered as a shortcut for the joining of two distinct points in the space-time or two distinct universe. The shape of the typical wormhole is like tube, which is asymptotically flat from both sides of the region. The
radius of the throat of the wormhole is either constant or variable depending on its structure and
the wormhole is termed as static or non-static respectively. One of the important essential condition to form a wormhole in general relativity is: the existence of an exotic matter component, which violates the energy conditions while the usual matter content satisfies these conditions. Subsequently, the general relativity predicts that the exotic form of matter component must be present near the throat of the wormhole\cite{Morris95}. Afterwards, it has been studied that exotic matter, which is threaded in the throat of wormhole, violates one of the energy condition, i.e., null energy condition(NEC)\cite{Visser96, Hochberg21}. Phantom energy is one possible candidate, which explains the cosmic accelerated expansion as well\cite{Jamil07, Jamil13, Lobo30}.
Moreover, the existence of phantom energy is problematic and no other appropriate exotic matter candidate is accessible, a substitute approach is usually monitored: the modifications of laws
of gravity, proposed primarily for elucidation of accelerated expansion and avoiding
singularities, could support the wormhole geometries. The presence of some form of energy- matter is necessary to sustain a wormhole solutions, because the wormhole is a non-vacuum solution of Einstein field equations. The matter content is assumed to satisfy the energy conditions near the throat of the wormhole, while the Lagrangian contains the higher order curvature terms which are required to sustain the wormhole solutions in modified gravity. Various authors studied wormhole geometries in modified gravity in different directions\cite{Lobo27, Lobo06, Lobo12, Garcia18, Sajadi53, Garcia018, Bertolami50, Harko04, Pavlovic17, Godani39, Samanta34, Samanta019, Godani019}.\\

\noindent
The general theory of relativity could be modified in various directions and this modification received huge consideration to explain accelerated expansion; wormhole structure; explaining flat rotation curves of galaxies and other mysterious phenomenon near black holes\cite{Nojiri15, Felice3, Capozziello67, Bamba55, Martino23}.\\

\noindent
The $f(R)$ modified gravity \cite{Buchdahl1} attains ample consideration for its capability to elucidate the accelerated expansion of the universe. In the early 1980s, Starobinsky \cite{Starobinsky99} discussed $f(R)$ model by taking $f(R)=R+\alpha R^2$, where $\alpha>0$, representing inflationary scenario of the universe. Simple idea on which this theory is:
the $f(R)$ theory of gravity replaces the scalar curvature $R$ in the Einstein gravitational
action to an arbitrary function $f(R)$, with $R$ being the leading order contribution to $f(R)$. Field equations obtained in this fashion have higher degree of complexity, and admit richer set of solutions than the standard general relativity.
A simplification of $f(R)$ gravity suggested in \cite{Bertolami16} integrates an unambiguous coupling between the matter Lagrangian and an arbitrary function of the scalar curvature, which leads to an extra force in the geodesic equation of a perfect fluid. Subsequently, it is shown that this extra force may be justification for the accelerated expansion of the universe \cite{Nojiri15, Bertolami46}.
The dynamical behavior of the matter and dark energy effects have been obtained within extended theories of gravity \cite{Nojiri001, Shirasaki, Capozziello001,  Rodrigues01}. Furthermore,  many authors have studied the dynamics of cosmological models in $f(R)$ gravity from various directions\cite{Capozziello91, Bombacigno21, Sbis, Chen, Elizalde26, Elizalde25, Astashenok1, Miranda, Nascimento, Odintsov20, Odintsov1, Nojiri56, Parth}.\\

\noindent
Apart from this some reviews on wormhole modeling in $f(R)$ gravity are given as follows:
Lobo and Oliveira \cite{Lobo12} constructed traversable wormhole models in modified $f(R)$ gravity and, they concluded that the higher order curvature derivative terms are responsible for the violation of the null energy condition and supporting the nonstandard wormhole structures. They considered constant redshift function, some specific shape functions and various equations of state to find the exact solutions.
%Cataldo et al. \cite{cata} investigated Lorentzian wormhole solutions for Einstien's field equations. They used barotropic equation of state for radial and lateral pressures and explored static and evolving wormholes in $N + 1$ dimensions.
Saiedi and Esfahani \cite{saiedi} considered shape and redshift functions as constant and scale factor as some positive power of cosmic time. They investigated wormhole solutions in $f(R)$ gravity and examined null and weak energy conditions.
Bouhmadi-L\'{o}pez et al. \cite{lopez} considered the sum of energy density and radial pressure to be proportional to a constant less than the area of the wormhole mouth. They inspected the solutions of spherically symmetric wormhole and analyze the stability regions.
Najafi et al. \cite{najafi} took an extra space-like dimension and studied traversable wormhole in FLRW model. They analyzed the effect of extra dimension on energy density, scale factor and shape function.
Bahamonde et al. \cite{baha} studied cosmological wormhole in $f(R)$ theory of gravity. They built a dynamical wormhole asymptotically approaching towards the FLRW universe and used the approximation of small wormholes for analysis. For the wormholes they considered, it was found that the presence of exotic matter near the throat is not needed, however it is always needed in case of general relativity.
Rahaman et al. \cite{Rahaman} studied wormhole solutions in Finslerian structure of space-time. They presented a wide variety of solutions and explored wormhole geometry by considering different choices of shape function and energy density.
%2017
Zubair et al. \cite{zubair} investigated wormhole solutions in the context of generalized $f(R, \phi)$ gravity for three types of fluids. They explored energy conditions and obtained wormhole solutions without need of exotic matter.
Kuhfittig \cite{peter} considered non-commutative geometry and discussed the existence of wormholes in $f(R)$ gravity. He considered various shape functions and obtained wormhole solutions satisfying general properties. He also considered $f(R) = \alpha R^2$ and determined wormhole solutions.
Novikov \cite{novi} reviewed wormholes and categorize them into three classes. They determined the properties of wormholes and described the  relation between black holes and wormholes. Sajadi and Riazi\cite{Sajadi18} obtained  static multi-polytropic wormhole solutions in the framework of general relativity, and they examine gravitational lensing by the wormhole, and calculate the deflection angle for weak and strong field limits as well.
Subsequently, various authors have been studied wormholes in different contexts \cite{Zangeneh1, Mehdizadeha, Zangeneh, Moraes, Bejarano, Mehdizadeh, Rogatko, Paul, Shaikh, Ovgun, Tsukamoto}.
%2018
Recently, Barros and Lobo \cite{barros} used three form fields and studied  static and spherically symmetric wormhole structures. They found various numerical and analytical solutions and showed that in the presence of three-form fields the null and weak energy conditions are satisfied in whole space-time. Godani and Samanta\cite{Godani39} and Samanta et al \cite{Samanta34} investigated wormhole solutions by defining new form function in $f(R)$ gravity and they tried to show the minimum requirement of exotic matter near the throat of the wormhole.\\

\noindent
The motivation of this paper is to study the wormhole solutions in viable $f(R)$ gravity by assuming some specific form of shape function. Subsequently, we have derived $f(R)$ function from the wormhole equations, which should satisfy the viability condition of $f(R)$ gravity. Furthermore, we have tried to avoid the presence of exotic matter near the throat by defining the suitable range of wormhole throat, so that energy condition could satisfy near the throat of the wormhole.
%Recent observations of supernovae of type Ia (SNe Ia) \cite{Grant, Perlmutter} and cosmic microwave background radiation  \cite{Bennett, Hinshaw} have suggested the cosmological expansion of universe in an accelerating phase. In literature, several candidate theories are introduced to describe this accelerated expansion.
\section{Traversable Wormhole \& Equations}
%In this section, $f(R)$ gravity and Einstein's field equations for FRW metric are described  briefly. Throughout dot and dash upon a function denote derivative with respect to cosmic time and Ricci scalar respectively.
The  static and spherically symmetric metric for wormhole geometry  is given by
  \begin{equation}\label{metric}
ds^2=-e^{2\Phi(r)}dt^2+\frac{dr^2}{1-b(r)/r} + r^2(d\theta^2+\sin^2\theta^2\phi^2),
\end{equation}
where $r$ is the radial coordinate varying from $r_0\neq 0$ to $\infty$, $\Phi(r)$ is the redshift function and $b(r)$ is the shape function. The wormhole has a throat at its center that joins two  asymptotically flat space-times. The function $\Phi(r)$ is responsible for the determination of gravitational redshift, so we call this $\Phi(r)$ is a redshift function. For a traversable wormhole, event horizon should be absent and the effect of tidal gravitational forces should be very small on a traveler. Therefore, to avoid event horizon or non-traversable condition, we required $e^{2\Phi(r)}\ne 0$. And we may have $e^{2\Phi(r)}\to 0$ provided $\Phi(r)\to -\infty$, so to avoid this situation, we have to avoid $\Phi(r)\to -\infty$, so we required $\Phi(r)$ is every where finite.  Hence, accordingly, we will have to choose $\Phi(r)$ such a way that $e^{2\Phi(r)}\ne 0$. It is not  necessary that, $\Phi(r)$ is only constant, it could be variable as well. For asymptotically flat region $\Leftrightarrow $ we required $\Phi\to 0$ and no horizon or singularity $\Rightarrow $  $\Phi$ is every where finite\cite{Morris95}. However, in this paper, for simplicity we consider constant redshift function. The function $b(r)$ is responsible for the shape of wormhole that should  satisfy the  following conditions:
(i) $b(r_0)=r_0$, (ii) $\frac{b(r)-b'(r)r}{b(r)^2}>0$, (iii) $b'(r_0)-1\leq 0$, (iv) $\frac{b(r)}{r}<1$ for $r>r_0$ and (v) $\frac{b(r)}{r}\rightarrow 0$ as $r\rightarrow\infty$. 

%Morris \& Thorne \cite{morris95} defined traversable wormholes using  Einstein's general theory of relativity.
\noindent
The $f(R)$  theory of gravity,  a generalization of  Einstein's theory of relativity, generalizes the action as
\begin{equation}\label{action}
S_G=\dfrac{1}{2k}\int[f(R) + L_m]\sqrt{-g}d^4x,
\end{equation}
where $k=8\pi G$, $L_m$ and $g$ stand for the  matter Lagrangian density and  the  determinant of the metric $g_{\mu\nu}$ respectively. For simplicity $k$ is taken as unity.\\
\noindent
Differentiating Eq.(\ref{action}) with respect to the metric $g_{\mu\nu}$, the field equations are obtained as
\begin{equation}\label{fe}
FR_{\mu\nu} -\dfrac{1}{2}fg_{\mu\nu}-\triangledown_\mu\triangledown_\nu F+\square Fg_{\mu\nu}= T_{\mu\nu}^m,
\end{equation}	
where $R_{\mu\nu}$ and $R$ denote Ricci tensor and curvature scalar respectively and $F=\frac{df}{dR}$. The contraction of \ref{fe}, gives
\begin{equation}\label{trace}
FR-2f+3\square F=T,
\end{equation}
where $T=T^{\mu}_{\mu}$ is the trace of the stress energy tensor.

\noindent
From Eqs. \ref{fe} \& \ref{trace}, the effective field equation is obtained as
\begin{equation}
G_{\mu\nu}\equiv R_{\mu\nu}-\frac{1}{2}Rg_{\mu\nu}=T_{\mu\nu}^{eff},
\end{equation}
where $T_{\mu\nu}^{eff}=T_{\mu\nu}^{c}+T_{\mu\nu}^{m}/F$ and $T_{\mu\nu}^{c}=\frac{1}{F}[\triangledown_\mu\triangledown_\nu F-\frac{1}{4}g_{\mu\nu}(FR+\square F+T)]$.
The energy momentum tensor for the matter source of the wormholes is $T_{\mu\nu}=\frac{\partial L_m}{\partial g^{\mu\nu}}$, which is defined as
\begin{equation}
T_{\mu\nu} = (\rho + p_t)u_\mu u_\nu - p_tg_{\mu\nu}+(p_r-p_t)X_\mu X_\nu,
\end{equation}	
such that
\begin{equation}
u^{\mu}u_\mu=-1 \mbox{ and } X^{\mu}X_\mu=1,
\end{equation}
where $\rho$,  $p_t$ and $p_r$  stand for the energy density, tangential pressure and radial pressure respectively.

\noindent
The  Ricci scalar $R$ given by $R=\frac{2b'(r)}{r^2}$ and Einstein's field equations for the metric \ref{metric} in  $f(R)$ gravity are obtained as:
%\cite{lobo}:
%\begin{eqnarray}  \label{6}
\begin{equation}\label{e1}
\rho=\frac{Fb'(r)}{r^2}-H
\end{equation}
\begin{equation}\label{e2}
p_r=-\frac{b(r)F}{r^3}-\Bigg(1-\frac{b(r)}{r}\Bigg)\Bigg[F''+\frac{F'(rb'(r)-b(r))}{2r^2\Big(1-\frac{b(r)}{r}\Big)}\Bigg]+H
\end{equation}
\begin{equation}\label{e3}
p_t=\frac{F(b(r)-rb'(r))}{2r^3}-\frac{F'}{r}\Bigg(1-\frac{b(r)}{r}\Bigg)+H,
\end{equation}
where $H=\frac{1}{4}(FR+\square F+T)$ and prime upon a function denotes the derivative of that function with respect to  radial coordinate $r$.

\noindent
The anisotropy parameter is defined as
\begin{equation}
\triangle=p_t-p_r.
\end{equation}
The geometry is attractive or repulsive in nature according as $\triangle$ is negative or positive. If $\triangle = 0$, then the geometry has an isotropic pressure.

%The metric \ref{metric} is spherical and symmetric.

\section{Energy conditions}
The important energy conditions are the Null Energy Condition (NEC), Weak Energy Condition (WEC), Strong Energy Condition (SEC) and Dominant Energy Condition (DEC). For any null vector, the null energy condition (NEC) is defined as
$NEC\Leftrightarrow T_{\mu\nu}k^{\mu}k^{\nu}\geq 0$.
Alternately, in terms of the principal pressures NEC is defined as $NEC\Leftrightarrow ~~ \forall i, ~\rho+p_{i}\ge 0$.
For a timelike vector, the weak energy condition (WEC) is defined as $WEC\Leftrightarrow T_{\mu\nu}V^{\mu}V^{\nu}\ge 0$. In terms of the principal pressures, it is defined as $WEC\Leftrightarrow \rho\ge 0;$ and $\forall i,  ~~ \rho+p_{i}\ge 0$. For a timelike vector, the strong energy condition (SEC) is defined as $SEC\Leftrightarrow (T_{\mu\nu}-\frac{T}{2}g_{\mu\nu})V^{\mu}V^{\nu}\ge 0$, where $T$ is the trace of the stress-energy tensor. In terms of the principal pressures, SEC is defined as
$T=-\rho+\sum_j{p_j}$ and $SEC\Leftrightarrow \forall j, ~ \rho+p_j\geq 0, ~ \rho+\sum_j{p_j}\geq 0$.  For any timelike vector, the dominant energy condition (DEC) is defined as $DEC\Leftrightarrow T_{\mu\nu}V^{\mu}V^{\nu}\ge 0$, and
$T_{\mu\nu}V^{\mu}$ is not space like. In terms of the principal pressures $DEC\Leftrightarrow \rho\ge 0;$ and $\forall i, ~ p_i\in [-\rho, ~+\rho]$.\\

\noindent
 In this paper, these conditions are investigated in terms of principal pressures which are as follows:
\begin{itemize}
	\item [(I)] $\rho + p_r\geq 0$, $\rho + p_t\geq 0$ (NEC)
	\item [(II)] $\rho \geq 0$, $\rho + p_r\geq 0$, $\rho + p_t\geq 0$ (WEC)
	\item [(III)] $\rho + p_r\geq 0$, $\rho + p_t\geq 0$,  $\rho + p_r +2p_t\geq 0$ (SEC)
	\item [(IV)] $\rho\geq 0$, $\rho - \lvert p_r\rvert \geq 0$, $\rho - \lvert p_t\rvert \geq 0$ (DEC)
\end{itemize}
%NEC : $\rho + p_r\geq 0$, $\rho + p_t\geq 0$   \\
%WEC : $\rho \geq 0$, $\rho + p_r> 0$, $\rho + p_t> 0$   \\
%SEC : $\rho + p_r\geq 0$, $\rho + p_t\geq 0$,  $\rho + p_r +2p_t\geq 0$  \\
%DEC : $\rho - \lvert p_r\rvert \geq 0$, $\rho - \lvert p_t\rvert \geq 0$\\
\noindent
A normal matter always satisfies these energy conditions because it possesses positive pressure and positive energy density. The wormholes are non-vacuum solutions of Einstein's field equations and according to Einstein's field theory, they are filled with a matter which is different from the normal matter and is known as exotic matter. This matter does not validate the energy conditions.

\section{Wormhole Solutions \&  Viable $f(R)$ Model}
In this section, $f(R)$ model is derived and field equations mentioned in Section 2 are solved. Furthermore, the energy density $\rho$, energy condition terms $\rho+p_r$, $\rho+p_t$, $\rho+p_r+2p_t$, $\rho-|p_r|$, $\rho-|p_t|$ and anisotropy parameter $\triangle$ are computed.\\
\noindent
Several authors have studied wormhole modelling by considering power-law shape function\cite{Lobo30, Lobo12, Sajadi18, peter, Lobo11, Garattini15, Heydarzade23}. In this paper, we use
 $b(r)=r\big(\frac{r}{r_0}\big)^\gamma$, where $0<\gamma<1$, the equation of state $p_r=\omega\rho$, where $\omega$ is equation of state parameter  and Equations (\ref{e1}) \& (\ref{e2}), the function $f(R)$ is obtained  as
\begin{equation}\label{f}
f(R)=k ((\gamma+1) r (\omega+1)-(\gamma+1) \omega-1)^{\frac{\gamma^2-1}{\gamma+1}},
\end{equation}
where $k$ is constant of integration.
\noindent
For Simplicity, we have taken $k=1$. So, we have
\begin{equation}\label{f1}
f(R)= ((\gamma+1) r (\omega+1)-(\gamma+1) \omega-1)^{\frac{\gamma^2-1}{\gamma+1}},
\end{equation}
Using the Ricci scalar $R=\frac{2b^{'}(r)}{r^2}$ and the form function $b(r)=r\big(\frac{r}{r_0}\big)^\gamma$, we can write
\begin{equation}\label{}
  r=\left(\frac{r_0^{\gamma+1}}{2(\gamma+1)}\right)^{\frac{1}{\gamma-2}}R^{\frac{1}{\gamma-2}}
\end{equation}
Now, the equation \eqref{f1} becomes
\begin{equation}\label{frm}
  f(R)= \bigg[(\gamma+1)^{\frac{3-\gamma}{2-\gamma}}(\omega+1)r_0^{\frac{\gamma+1}{\gamma-2}}\frac{1}{2} R^{\frac{1}{\gamma-2}}-(\gamma+1) \omega-1\bigg]^{\gamma-1}
\end{equation}
In the early 1980s, Starobinsky\cite{Starobinsky99} showed that the model $f(R)=R+\alpha R^2$, where $\alpha>0$ can be responsible for inflationary phase of the early universe. The presence of quadratic term $\alpha R^2$ was responsible for this fact and gave rise to an asymptotically exact de Sitter solution. If the term $\alpha R^2$ becomes smaller than the linear term $R$, then the inflation will be stopped. Hence, this model is a suitable one to discuss the inflationary stage of the early universe. However, this model is not a suitable candidate to discuss the present cosmic accelerated expansion. Afterwards, the models $f(R)=R-\frac{\alpha}{R^n}$ (where $\alpha>0$ and $n>0$) were proposed as a candidate for dark energy to explain the late time cosmic accelerated expansion\cite{Capozziello83, Capozziello69, Carroll28, Nojiri12}. Moreover, because of the instability associated with negative value of $f_{,RR}$, these models are do not satisfy local gravity conditions\cite{Chiba1, Dolgov1, Soussa55, Olmo05, Faraoni29}. So this makes it very crucial to make a
set of conditions which are viable for $f(R)$ models in metric formalism. These conditions as stated as follows\cite{Amendola13}:
\begin{itemize}

  \item To avoid anti-gravity, we required, $f_{,R}>0$ for $R\ge R_0$, where $R_0$ is the Ricci scalar at present epoch and is positive.
  
  \item To avoid complex valued function of $f(R)$, we required $f(R)$ should be real valued function. For real valued condition of our derived $f(R)$ function, we required $\omega>-1.4$ and $r>1$.
  \item $f_{,RR}>0$ for $R\ge R_0$. This is required for consistency with local gravity tests\cite{Dolgov1, Olmo05, Faraoni29, Navarro22}, for the presence of the matter-dominated epoch\cite{Amendola04}, and for the stability of cosmological perturbations\cite{Carroll23, Song04, Bean20, Faulkner05}.

  \item $f(R)\to R-2\Lambda$ for $R\gg R_0$. This is required for consistency with local gravity tests\cite{Faulkner05, Hu04, Starobinsky, Appleby7, Tsujikawa07} and for the presence of the matter-dominated epoch\cite{Amendola04} .

\item $0<\frac{Rf_{,RR}}{f_{,R}}<1$ at $\frac{Rf_{,R}}{f}=2$. This is required for the stability of the late-time de Sitter point\cite{Amendola04,  Muller98, Faraoni37}.
\end{itemize}

\noindent
In this paper, the $f(R)$ model derived in equation \eqref{frm} is found to be satisfied all the above conditions. Therefore, we can say that our derived $f(R)$ function is a viable $f(R)$ model. Now, Using this viable $f(R)$ model and Equations \eqref{e1}, \eqref{e2} \& \eqref{e3}, we obtained the following terms:

\begin{eqnarray}
\rho&=&\frac{1}{4} \left(\frac{r}{{r_0}}\right)^{-\gamma} ((\gamma+1) r (\omega+1)-(\gamma+1) \omega-1)^{\gamma-4} \left[\gamma^4 (-\omega) (\omega+1) \left(r (\omega+1) \left(\frac{r}{{r_0}}\right)^\gamma\right.\right.\nonumber\\
&+&\left.\left.\omega \left(2-3 \left(\frac{r}{{r_0}}\right)^\gamma\right)\right)+\gamma^3 \left(2 r^2 (\omega+1)^2 \left(\frac{r}{{r_0}}\right)^\gamma-r \left(5 \omega^2+6 \omega+1\right) \left(\frac{r}{{r_0}}\right)^\gamma-2 \omega \left(3 \omega^2\right.\right.\right.\nonumber\\
&\times& \left.\left.\left.\left(\left(\frac{r}{{r_0}}\right)^\gamma-1\right)-\omega \left(\left(\frac{r}{{r_0}}\right)^\gamma+1\right)-3 \left(\frac{r}{{r_0}}\right)^\gamma+2\right)\right)+\gamma^2 (\omega+1) \left(4 r^2 (\omega+1) \left(\frac{r}{{r_0}}\right)^\gamma\right.\right.\nonumber\\
&+&\left.\left.r \left(3 \omega^2-4 \omega-3\right) \left(\frac{r}{{r_0}}\right)^\gamma+\omega^2 \left(6-9 \left(\frac{r}{{r_0}}\right)^\gamma\right)-2 \omega \left(7 \left(\frac{r}{{r_0}}\right)^\gamma-8\right)+3 \left(\frac{r}{{r_0}}\right)^\gamma-2\right)\right.\nonumber\\
&+&\left.2 \gamma (\omega+1)^2 \left(r^2 \left(\frac{r}{{r_0}}\right)^\gamma+r (\omega-1) \left(\frac{r}{{r_0}}\right)^\gamma+\omega \left(6 \left(\frac{r}{{r_0}}\right)^\gamma-7\right)-5 \left(\frac{r}{{r_0}}\right)^\gamma+5\right)\right.\nonumber\\
&+&\left.12 (\omega+1)^3 \left(\left(\frac{r}{{r_0}}\right)^\gamma-1\right)\right]\nonumber
\end{eqnarray}
\noindent
The energy density $\rho$ is plotted in figure(a) with respect to the radial coordinate $r$, and it is obtained that the energy density is positive for $0.43\leq\gamma<1$, $\omega\geq 0$ and $r\geq0.9$. From this computation, we realize, if we consider the size of the throat of the wormhole is either $r_0=0.9$ or may be $r_0>0.9$, then we do not observe any negative energy density. That means, the throat of the wormhole is filled with non-exotic matter.
\begin{eqnarray}
p_r&=&\frac{1}{4}\omega \left(\frac{r}{{r_0}}\right)^{-\gamma} ((\gamma+1) r (\omega+1)-(\gamma+1) \omega-1)^{\gamma-4} \left[\gamma^4 (-\omega) (\omega+1) \left(r (\omega+1) \left(\frac{r}{{r_0}}\right)^\gamma\right.\right.\nonumber\\
&+&\left.\left.\omega \left(2-3 \left(\frac{r}{{r_0}}\right)^\gamma\right)\right)+\gamma^3 \left(2 r^2 (\omega+1)^2 \left(\frac{r}{{r_0}}\right)^\gamma-r \left(5 \omega^2+6 \omega+1\right) \left(\frac{r}{{r_0}}\right)^\gamma-2 \omega \left(3 \omega^2\right.\right.\right.\nonumber\\
&\times& \left.\left.\left.\left(\left(\frac{r}{{r_0}}\right)^\gamma-1\right)-\omega \left(\left(\frac{r}{{r_0}}\right)^\gamma+1\right)-3 \left(\frac{r}{{r_0}}\right)^\gamma+2\right)\right)+\gamma^2 (\omega+1) \left(4 r^2 (\omega+1) \left(\frac{r}{{r_0}}\right)^\gamma\right.\right.\nonumber\\
&+&\left.\left.r \left(3 \omega^2-4 \omega-3\right) \left(\frac{r}{{r_0}}\right)^\gamma+\omega^2 \left(6-9 \left(\frac{r}{{r_0}}\right)^\gamma\right)-2 \omega \left(7 \left(\frac{r}{{r_0}}\right)^\gamma-8\right)+3 \left(\frac{r}{{r_0}}\right)^\gamma-2\right)\right.\nonumber\\
&+&\left.2 \gamma (\omega+1)^2 \left(r^2 \left(\frac{r}{{r_0}}\right)^\gamma+r (\omega-1) \left(\frac{r}{{r_0}}\right)^\gamma+\omega \left(6 \left(\frac{r}{{r_0}}\right)^\gamma-7\right)-5 \left(\frac{r}{{r_0}}\right)^\gamma+5\right)\right.\nonumber\\
&+&\left.12 (\omega+1)^3 \left(\left(\frac{r}{{r_0}}\right)^\gamma-1\right)\right]\nonumber
\end{eqnarray}

\begin{eqnarray}
p_t&=&\frac{1}{4} \left(\frac{r}{{r_0}}\right)^{-\gamma} ((\gamma+1) r (\omega+1)-(\gamma+1) \omega-1)^{\gamma-4} \left(\gamma^4 \omega (\omega+1) \left(r (\omega+1) \left(\frac{r}{{r_0}}\right)^\gamma\right.\right.\nonumber\\
&+&\left.\left.\omega \left(2-3 \left(\frac{r}{{r_0}}\right)^\gamma\right)\right)+\gamma^3 \left(r^2 (\omega-1) (\omega+1)^2 \left(\frac{r}{{r_0}}\right)^\gamma+\omega \left(\omega^2 \left(5 \left(\frac{r}{{r_0}}\right)^\gamma-4\right)\right.\right.\right.\nonumber\\
&-&\left.\left.\left.3 \omega \left(\frac{r}{{r_0}}\right)^\gamma-6 \left(\frac{r}{{r_0}}\right)^\gamma+4\right)+r (\omega+1) \left(\omega \left(5 \left(\frac{r}{{r_0}}\right)^\gamma-2\right)+\left(\frac{r}{{r_0}}\right)^\gamma-2 \omega^2\right)\right)\right.\nonumber\\
&+&\left.\gamma^2 (\omega+1) \left(4 r^2 \omega (\omega+1) \left(\frac{r}{{r_0}}\right)^\gamma+\omega^2 \left(13 \left(\frac{r}{{r_0}}\right)^\gamma-6\right)-r \left(11 \omega^2 \left(\frac{r}{{r_0}}\right)^\gamma+\right.\right.\right.\nonumber
\end{eqnarray}
\begin{eqnarray}
&+&\left.\left.\left.\left.\omega \left(4 \left(\frac{r}{{r_0}}\right)^\gamma+52\right)-3 \left(\frac{r}{{r_0}}\right)^\gamma+2\right)+12 \omega \left(\left(\frac{r}{{r_0}}\right)^\gamma-1\right)-3 \left(\frac{r}{{r_0}}\right)^\gamma+2\right) \right.\right.cv\nonumber\\
&&\left.\gamma (\omega+1)^2 \left(r^2 (5 \omega+3)\left(\frac{r}{{r_0}}\right)^\gamma-\omega \left(\left(\frac{r}{{r_0}}\right)^\gamma-8\right)-2 r (3 \omega+1) \left(3 \left(\frac{r}{{r_0}}\right)^\gamma-1\right)\right.\right.\nonumber\\
&+&\left.\left.\left.+9 \left(\frac{r}{{r_0}}\right)^\gamma-8\right)+2 (\omega+1)^3 \left(r^2 \left(\frac{r}{{r_0}}\right)^\gamma\right.-3 \left(\frac{r}{{r_0}}\right)^\gamma+r \left(2-4 \left(\frac{r}{{r_0}}\right)^\gamma\right)+4\right)\right)
\end{eqnarray}

\begin{eqnarray}
\rho+p_r&=&\frac{1}{4}(1+\omega) \left(\frac{r}{{r_0}}\right)^{-\gamma} ((\gamma+1) r (\omega+1)-(\gamma+1) \omega-1)^{\gamma-4} \left[\gamma^4 (-\omega) (\omega+1) \left(r (\omega+1) \left(\frac{r}{{r_0}}\right)^\gamma\right.\right.\nonumber\\
&+&\left.\left.\omega \left(2-3 \left(\frac{r}{{r_0}}\right)^\gamma\right)\right)+\gamma^3 \left(2 r^2 (\omega+1)^2 \left(\frac{r}{{r_0}}\right)^\gamma-r \left(5 \omega^2+6 \omega+1\right) \left(\frac{r}{{r_0}}\right)^\gamma-2 \omega \left(3 \omega^2\right.\right.\right.\nonumber\\
&\times& \left.\left.\left.\left(\left(\frac{r}{{r_0}}\right)^\gamma-1\right)-\omega \left(\left(\frac{r}{{r_0}}\right)^\gamma+1\right)-3 \left(\frac{r}{{r_0}}\right)^\gamma+2\right)\right)+\gamma^2 (\omega+1) \left(4 r^2 (\omega+1) \left(\frac{r}{{r_0}}\right)^\gamma\right.\right.\nonumber\\
&+&\left.\left.r \left(3 \omega^2-4 \omega-3\right) \left(\frac{r}{{r_0}}\right)^\gamma+\omega^2 \left(6-9 \left(\frac{r}{{r_0}}\right)^\gamma\right)-2 \omega \left(7 \left(\frac{r}{{r_0}}\right)^\gamma-8\right)+3 \left(\frac{r}{{r_0}}\right)^\gamma-2\right)\right.\nonumber\\
&+&\left.2 \gamma (\omega+1)^2 \left(r^2 \left(\frac{r}{{r_0}}\right)^\gamma+r (\omega-1) \left(\frac{r}{{r_0}}\right)^\gamma+\omega \left(6 \left(\frac{r}{{r_0}}\right)^\gamma-7\right)-5 \left(\frac{r}{{r_0}}\right)^\gamma+5\right)\right.\nonumber\\
&+&\left.12 (\omega+1)^3 \left(\left(\frac{r}{{r_0}}\right)^\gamma-1\right)\right]\nonumber
\end{eqnarray}
\noindent
In figure(b), the behavior of the term $\rho+p_r$ is plotted and it is found that the term $\rho+p_r$ is positive for $0.43\leq\gamma<1$, $\omega\geq 0$ and $r\geq0.9$.
\begin{eqnarray}
\rho+p_t&=&\frac{1}{4} \left(\frac{r}{{r_0}}\right)^{-\gamma} ((\gamma+1) r (\omega+1)-(\gamma+1) \omega-1)^{\gamma-4} \left[\gamma^4 (-\omega) (\omega+1) \left(r (\omega+1) \left(\frac{r}{{r_0}}\right)^\gamma\right.\right.\nonumber\\
&+&\left.\left.\omega \left(2-3 \left(\frac{r}{{r_0}}\right)^\gamma\right)\right)+\gamma^3 \left(2 r^2 (\omega+1)^2 \left(\frac{r}{{r_0}}\right)^\gamma-r \left(5 \omega^2+6 \omega+1\right) \left(\frac{r}{{r_0}}\right)^\gamma-2 \omega \left(3 \omega^2\right.\right.\right.\nonumber\\
&\times& \left.\left.\left.\left(\left(\frac{r}{{r_0}}\right)^\gamma-1\right)-\omega \left(\left(\frac{r}{{r_0}}\right)^\gamma+1\right)-3 \left(\frac{r}{{r_0}}\right)^\gamma+2\right)\right)+\gamma^2 (\omega+1) \left(4 r^2 (\omega+1) \left(\frac{r}{{r_0}}\right)^\gamma\right.\right.\nonumber\\
&+&\left.\left.r \left(3 \omega^2-4 \omega-3\right) \left(\frac{r}{{r_0}}\right)^\gamma+\omega^2 \left(6-9 \left(\frac{r}{{r_0}}\right)^\gamma\right)-2 \omega \left(7 \left(\frac{r}{{r_0}}\right)^\gamma-8\right)+3 \left(\frac{r}{{r_0}}\right)^\gamma-2\right)\right.\nonumber\\
&+&\left.2 \gamma (\omega+1)^2 \left(r^2 \left(\frac{r}{{r_0}}\right)^\gamma+r (\omega-1) \left(\frac{r}{{r_0}}\right)^\gamma+\omega \left(6 \left(\frac{r}{{r_0}}\right)^\gamma-7\right)-5 \left(\frac{r}{{r_0}}\right)^\gamma+5\right)\right.\nonumber\\
&+&\left.12 (\omega+1)^3 \left(\left(\frac{r}{{r_0}}\right)^\gamma-1\right)\right]\nonumber\\
&+&\frac{1}{4} \left(\frac{r}{{r_0}}\right)^{-\gamma} ((\gamma+1) r (\omega+1)-(\gamma+1) \omega-1)^{\gamma-4} \left(\gamma^4 \omega (\omega+1) \left(r (\omega+1) \left(\frac{r}{{r_0}}\right)^\gamma\right.\right.\nonumber\\
&+&\left.\left.\omega \left(2-3 \left(\frac{r}{{r_0}}\right)^\gamma\right)\right)+\gamma^3 \left(r^2 (\omega-1) (\omega+1)^2 \left(\frac{r}{{r_0}}\right)^\gamma+\omega \left(\omega^2 \left(5 \left(\frac{r}{{r_0}}\right)^\gamma-4\right)\right.\right.\right.\nonumber
\end{eqnarray}

\begin{eqnarray}
&-&\left.\left.\left.3 \omega \left(\frac{r}{{r_0}}\right)^\gamma-6 \left(\frac{r}{{r_0}}\right)^\gamma+4\right)+r (\omega+1) \left(\omega \left(5 \left(\frac{r}{{r_0}}\right)^\gamma-2\right)+\left(\frac{r}{{r_0}}\right)^\gamma-2 \omega^2\right)\right)\right.\nonumber\\
&+&\left.\gamma^2 (\omega+1) \left(4 r^2 \omega (\omega+1) \left(\frac{r}{{r_0}}\right)^\gamma+\omega^2 \left(13 \left(\frac{r}{{r_0}}\right)^\gamma-6\right)-r \left(11 \omega^2 \left(\frac{r}{{r_0}}\right)^\gamma+\omega \left(4 \left(\frac{r}{{r_0}}\right)^\gamma\right.\right.\right.\right.\nonumber\\
&+&\left.\left.\left.\left.52\right)-3 \left(\frac{r}{{r_0}}\right)^\gamma+2\right)+12 \omega \left(\left(\frac{r}{{r_0}}\right)^\gamma-1\right)-3 \left(\frac{r}{{r_0}}\right)^\gamma+2\right)+\gamma (\omega+1)^2 \left(r^2 (5 \omega+3) \right.\right.\nonumber\\
&\times&\left.\left.\left(\frac{r}{{r_0}}\right)^\gamma-\omega \left(\left(\frac{r}{{r_0}}\right)^\gamma-8\right)-2 r (3 \omega+1) \left(3 \left(\frac{r}{{r_0}}\right)^\gamma-1\right)+9 \left(\frac{r}{{r_0}}\right)^\gamma-8\right)+2 (\omega+1)^3 \right.\nonumber\\
&\times&\left.\left.\left(r^2 \left(\frac{r}{{r_0}}\right)^\gamma\right.-3 \left(\frac{r}{{r_0}}\right)^\gamma+r \left(2-4 \left(\frac{r}{{r_0}}\right)^\gamma\right)+4\right)\right)
\end{eqnarray}
\noindent
In figure(c), the behavior of the term $\rho+p_t$ is plotted and it is found that the term $\rho+p_t$ is positive for $0.43\leq\gamma<1$, $\omega\geq 0$ and $r\geq 1.4$.
\begin{eqnarray}
\rho+p_r+2p_t&=&\frac{1}{4}(1+\omega) \left(\frac{r}{{r_0}}\right)^{-\gamma} ((\gamma+1) r (\omega+1)-(\gamma+1) \omega-1)^{\gamma-4} \left[\gamma^4 (-\omega) (\omega+1) \left(r (\omega+1) \right.\right.\nonumber\\
&\times&\left.\left.\left(\frac{r}{{r_0}}\right)^\gamma+\omega \left(2-3 \left(\frac{r}{{r_0}}\right)^\gamma\right)\right)+\gamma^3 \left(2 r^2 (\omega+1)^2 \left(\frac{r}{{r_0}}\right)^\gamma-r \left(5 \omega^2+6 \omega+1\right) \left(\frac{r}{{r_0}}\right)^\gamma\right.\right.\nonumber\\
&-& \left.\left.2 \omega \left(3 \omega^2\left(\left(\frac{r}{{r_0}}\right)^\gamma-1\right)-\omega \left(\left(\frac{r}{{r_0}}\right)^\gamma+1\right)-3 \left(\frac{r}{{r_0}}\right)^\gamma+2\right)\right)+\gamma^2 (\omega+1) \left(4 r^2 \right.\right.\nonumber\\
&\times&\left.\left.(\omega+1) \left(\frac{r}{{r_0}}\right)^\gamma+r \left(3 \omega^2-4 \omega-3\right) \left(\frac{r}{{r_0}}\right)^\gamma+\omega^2 \left(6-9 \left(\frac{r}{{r_0}}\right)^\gamma\right)-2 \omega \left(7 \left(\frac{r}{{r_0}}\right)^\gamma\right.\right.\right.\nonumber\\
&-&\left.\left.\left.8\right)+3 \left(\frac{r}{{r_0}}\right)^\gamma-2\right)+2 \gamma (\omega+1)^2 \left(r^2 \left(\frac{r}{{r_0}}\right)^\gamma+r (\omega-1) \left(\frac{r}{{r_0}}\right)^\gamma+\omega \left(6 \left(\frac{r}{{r_0}}\right)^\gamma
\right.\right.\right.\nonumber
\\
&-&\left.\left.\left.\left.\left.7\right)-5 \left(\frac{r}{{r_0}}\right)^\gamma+5\right)8\right)+3 \left(\frac{r}{{r_0}}\right)^\gamma-2\right)+12 (\omega+1)^3 \left(\left(\frac{r}{{r_0}}\right)^\gamma-1\right)\right]\nonumber\\
&+&\frac{1}{2} \left(\frac{r}{{r_0}}\right)^{-\gamma} ((\gamma+1) r (\omega+1)-(\gamma+1) \omega-1)^{\gamma-4} \left(\gamma^4 \omega (\omega+1) \left(r (\omega+1) \left(\frac{r}{{r_0}}\right)^\gamma\right.\right.\nonumber\\
&+&\left.\left.\omega \left(2-3 \left(\frac{r}{{r_0}}\right)^\gamma\right)\right)+\gamma^3 \left(r^2 (\omega-1) (\omega+1)^2 \left(\frac{r}{{r_0}}\right)^\gamma+\omega \left(\omega^2 \left(5 \left(\frac{r}{{r_0}}\right)^\gamma-4\right)\right.\right.\right.\nonumber\\
&-&\left.\left.\left.3 \omega \left(\frac{r}{{r_0}}\right)^\gamma-6 \left(\frac{r}{{r_0}}\right)^\gamma+4\right)+r (\omega+1) \left(\omega \left(5 \left(\frac{r}{{r_0}}\right)^\gamma-2\right)+\left(\frac{r}{{r_0}}\right)^\gamma-2 \omega^2\right)\right)\right.\nonumber\\
&+&\left.\gamma^2 (\omega+1) \left(4 r^2 \omega (\omega+1) \left(\frac{r}{{r_0}}\right)^\gamma+\omega^2 \left(13 \left(\frac{r}{{r_0}}\right)^\gamma-6\right)-r \left(11 \omega^2 \left(\frac{r}{{r_0}}\right)^\gamma\right.\right.\right.\nonumber\\
&+&\left.\left.\left.+\omega \left(4 \left(\frac{r}{{r_0}}\right)^\gamma+52\right)-3 \left(\frac{r}{{r_0}}\right)^\gamma+2\right)+12 \omega \left(\left(\frac{r}{{r_0}}\right)^\gamma-1\right)-3 \left(\frac{r}{{r_0}}\right)^\gamma+2\right) \right.\nonumber\\
&+&\gamma (\omega+1)^2 \left(r^2 (5 \omega+3)\left(\frac{r}{{r_0}}\right)^\gamma-\omega \left(\left(\frac{r}{{r_0}}\right)^\gamma-8\right)-2 r (3 \omega+1) \left(3 \left(\frac{r}{{r_0}}\right)^\gamma-1\right)\right. \nonumber\\
&+&\left.\left.9 \left(\frac{r}{{r_0}}\right)^\gamma-8\right)+2 (\omega+1)^3\left(r^2 \left(\frac{r}{{r_0}}\right)^\gamma-3 \left(\frac{r}{{r_0}}\right)^\gamma+r \left(2-4 \left(\frac{r}{{r_0}}\right)^\gamma\right)+4\right)\right)
\end{eqnarray}
\noindent
The behavior of the term $\rho+p_r+2p_t$ is plotted in the figure(d) with respect the radial coordinate $r$, and it is found to be positive for $0.7\leq\gamma<1$, $\omega\geq 0$ and $r\geq1.7$.
\begin{eqnarray}
\rho-|p_r|&=&\frac{1}{4} \left(\frac{r}{{r_0}}\right)^{-\gamma} ((\gamma+1) r (\omega+1)-(\gamma+1) \omega-1)^{\gamma-4} \left[\gamma^4 (-\omega) (\omega+1) \left(r (\omega+1) \left(\frac{r}{{r_0}}\right)^\gamma\right.\right.\nonumber\\
&+&\left.\left.\omega \left(2-3 \left(\frac{r}{{r_0}}\right)^\gamma\right)\right)+\gamma^3 \left(2 r^2 (\omega+1)^2 \left(\frac{r}{{r_0}}\right)^\gamma-r \left(5 \omega^2+6 \omega+1\right) \left(\frac{r}{{r_0}}\right)^\gamma-2 \omega \left(3 \omega^2\right.\right.\right.\nonumber\\
&\times& \left.\left.\left.\left(\left(\frac{r}{{r_0}}\right)^\gamma-1\right)-\omega \left(\left(\frac{r}{{r_0}}\right)^\gamma+1\right)-3 \left(\frac{r}{{r_0}}\right)^\gamma+2\right)\right)+\gamma^2 (\omega+1) \left(4 r^2 (\omega+1) \left(\frac{r}{{r_0}}\right)^\gamma\right.\right.\nonumber\\
&+&\left.\left.r \left(3 \omega^2-4 \omega-3\right) \left(\frac{r}{{r_0}}\right)^\gamma+\omega^2 \left(6-9 \left(\frac{r}{{r_0}}\right)^\gamma\right)-2 \omega \left(7 \left(\frac{r}{{r_0}}\right)^\gamma-8\right)+3 \left(\frac{r}{{r_0}}\right)^\gamma-2\right)\right.\nonumber\\
&+&\left.2 \gamma (\omega+1)^2 \left(r^2 \left(\frac{r}{{r_0}}\right)^\gamma+r (\omega-1) \left(\frac{r}{{r_0}}\right)^\gamma+\omega \left(6 \left(\frac{r}{{r_0}}\right)^\gamma-7\right)-5 \left(\frac{r}{{r_0}}\right)^\gamma+5\right)\right.\nonumber\\
&+&\left.12 (\omega+1)^3 \left(\left(\frac{r}{{r_0}}\right)^\gamma-1\right)\right]\nonumber\\
&-&\Bigg|\frac{1}{4}\omega \left(\frac{r}{{r_0}}\right)^{-\gamma} ((\gamma+1) r (\omega+1)-(\gamma+1) \omega-1)^{\gamma-4} \left[\gamma^4 (-\omega) (\omega+1) \left(r (\omega+1) \left(\frac{r}{{r_0}}\right)^\gamma\right.\right.\nonumber\\
&+&\left.\left.\omega \left(2-3 \left(\frac{r}{{r_0}}\right)^\gamma\right)\right)+\gamma^3 \left(2 r^2 (\omega+1)^2 \left(\frac{r}{{r_0}}\right)^\gamma-r \left(5 \omega^2+6 \omega+1\right) \left(\frac{r}{{r_0}}\right)^\gamma-2 \omega \left(3 \omega^2\right.\right.\right.\nonumber\\
&\times& \left.\left.\left.\left(\left(\frac{r}{{r_0}}\right)^\gamma-1\right)-\omega \left(\left(\frac{r}{{r_0}}\right)^\gamma+1\right)-3 \left(\frac{r}{{r_0}}\right)^\gamma+2\right)\right)+\gamma^2 (\omega+1) \left(4 r^2 (\omega+1) \left(\frac{r}{{r_0}}\right)^\gamma\right.\right.\nonumber\\
&+&\left.\left.r \left(3 \omega^2-4 \omega-3\right) \left(\frac{r}{{r_0}}\right)^\gamma+\omega^2 \left(6-9 \left(\frac{r}{{r_0}}\right)^\gamma\right)-2 \omega \left(7 \left(\frac{r}{{r_0}}\right)^\gamma-8\right)+3 \left(\frac{r}{{r_0}}\right)^\gamma-2\right)\right.\nonumber\\
&+&\left.2 \gamma (\omega+1)^2 \left(r^2 \left(\frac{r}{{r_0}}\right)^\gamma+r (\omega-1) \left(\frac{r}{{r_0}}\right)^\gamma+\omega \left(6 \left(\frac{r}{{r_0}}\right)^\gamma-7\right)-5 \left(\frac{r}{{r_0}}\right)^\gamma+5\right)\right.\nonumber\\
&+&\left.12 (\omega+1)^3 \left(\left(\frac{r}{{r_0}}\right)^\gamma-1\right)\right]\Bigg|\nonumber
\end{eqnarray}

\noindent
The nature of the dominate energy condition term $\rho-|p_r|$ is plotted in figure(e) and it is found to be positive only for $0.2\leq\gamma<1$, $0\leq\omega\leq 0.9$ and $r\geq1.6$.

\begin{eqnarray}
\rho-|p_t|&=&\frac{1}{4} \left(\frac{r}{{r_0}}\right)^{-\gamma} ((\gamma+1) r (\omega+1)-(\gamma+1) \omega-1)^{\gamma-4} \left[\gamma^4 (-\omega) (\omega+1) \left(r (\omega+1) \left(\frac{r}{{r_0}}\right)^\gamma\right.\right.\nonumber\\
&+&\left.\left.\omega \left(2-3 \left(\frac{r}{{r_0}}\right)^\gamma\right)\right)+\gamma^3 \left(2 r^2 (\omega+1)^2 \left(\frac{r}{{r_0}}\right)^\gamma-r \left(5 \omega^2+6 \omega+1\right) \left(\frac{r}{{r_0}}\right)^\gamma-2 \omega \left(3 \omega^2\right.\right.\right.\nonumber\\
&\times& \left.\left.\left.\left(\left(\frac{r}{{r_0}}\right)^\gamma-1\right)-\omega \left(\left(\frac{r}{{r_0}}\right)^\gamma+1\right)-3 \left(\frac{r}{{r_0}}\right)^\gamma+2\right)\right)+\gamma^2 (\omega+1) \left(4 r^2 (\omega+1) \left(\frac{r}{{r_0}}\right)^\gamma\right.\right.\nonumber\\
&+&\left.\left.r \left(3 \omega^2-4 \omega-3\right) \left(\frac{r}{{r_0}}\right)^\gamma+\omega^2 \left(6-9 \left(\frac{r}{{r_0}}\right)^\gamma\right)-2 \omega \left(7 \left(\frac{r}{{r_0}}\right)^\gamma-8\right)+3 \left(\frac{r}{{r_0}}\right)^\gamma-2\right)\right.\nonumber\\
&+&\left.2 \gamma (\omega+1)^2 \left(r^2 \left(\frac{r}{{r_0}}\right)^\gamma+r (\omega-1) \left(\frac{r}{{r_0}}\right)^\gamma+\omega \left(6 \left(\frac{r}{{r_0}}\right)^\gamma-7\right)-5 \left(\frac{r}{{r_0}}\right)^\gamma+5\right)\right.\nonumber\\
&+&\left.12 (\omega+1)^3 \left(\left(\frac{r}{{r_0}}\right)^\gamma-1\right)\right]\nonumber
\end{eqnarray}
\begin{eqnarray}
&-&\Bigg|\frac{1}{4} \left(\frac{r}{{r_0}}\right)^{-\gamma} ((\gamma+1) r (\omega+1)-(\gamma+1) \omega-1)^{\gamma-4} \left(\gamma^4 \omega (\omega+1) \left(r (\omega+1) \left(\frac{r}{{r_0}}\right)^\gamma\right.\right.\nonumber\\
&+&\left.\left.\omega \left(2-3 \left(\frac{r}{{r_0}}\right)^\gamma\right)\right)+\gamma^3 \left(r^2 (\omega-1) (\omega+1)^2 \left(\frac{r}{{r_0}}\right)^\gamma+\omega \left(\omega^2 \left(5 \left(\frac{r}{{r_0}}\right)^\gamma-4\right)\right.\right.\right.\nonumber\\
&-&\left.\left.\left.3 \omega \left(\frac{r}{{r_0}}\right)^\gamma-6 \left(\frac{r}{{r_0}}\right)^\gamma+4\right)+r (\omega+1) \left(\omega \left(5 \left(\frac{r}{{r_0}}\right)^\gamma-2\right)+\left(\frac{r}{{r_0}}\right)^\gamma-2 \omega^2\right)\right)\right.\nonumber\\
&+&\left.\gamma^2 (\omega+1) \left(4 r^2 \omega (\omega+1) \left(\frac{r}{{r_0}}\right)^\gamma+\omega^2 \left(13 \left(\frac{r}{{r_0}}\right)^\gamma-6\right)-r \left(11 \omega^2 \left(\frac{r}{{r_0}}\right)^\gamma+\omega \left(4 \left(\frac{r}{{r_0}}\right)^\gamma\right.\right.\right.\right.\nonumber\\
&+&\left.\left.\left.\left.52\right)-3 \left(\frac{r}{{r_0}}\right)^\gamma+2\right)+12 \omega \left(\left(\frac{r}{{r_0}}\right)^\gamma-1\right)-3 \left(\frac{r}{{r_0}}\right)^\gamma+2\right)+\gamma (\omega+1)^2 \left(r^2 (5 \omega+3) \right.\right.\nonumber\\
&\times&\left.\left.\left(\frac{r}{{r_0}}\right)^\gamma-\omega \left(\left(\frac{r}{{r_0}}\right)^\gamma-8\right)-2 r (3 \omega+1) \left(3 \left(\frac{r}{{r_0}}\right)^\gamma-1\right)+9 \left(\frac{r}{{r_0}}\right)^\gamma-8\right)+2 (\omega+1)^3 \right.\nonumber\\
&\times&\left.\left.\left(r^2 \left(\frac{r}{{r_0}}\right)^\gamma\right.-3 \left(\frac{r}{{r_0}}\right)^\gamma+r \left(2-4 \left(\frac{r}{{r_0}}\right)^\gamma\right)+4\right)\right)\Bigg|
\end{eqnarray}

\noindent
The nature of the second dominate energy condition term $\rho-|p_t|$ is plotted in figure(f) and it is found to be positive only for $0.2\leq\gamma<1$, $0\leq\omega\leq 0.9$ and $r\geq1.6$.

\begin{figure}
	\centering
	\subfigure[The energy density $\rho$ is positive for $0.43\leq\gamma<1$, $\omega\geq 0$ and $r\geq0.9$. So in this figure we plotted the energy density with respect to $r$ and $\omega$ taking $\gamma = 0.5$.  ]{\includegraphics[scale=.65]{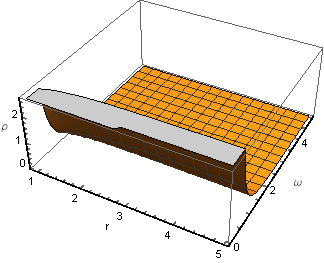}}\hspace{.1cm}
	\subfigure[The first NEC term $\rho+p_r$ is obtained to be positive for $0.43\leq\gamma<1$, $\omega\geq 0$ and $r\geq0.9$, so in this figure we plotted $\rho+p_r$ with respect to $r$ for $\gamma = 0.5$ and $\omega>0$.]{\includegraphics[scale=.65]{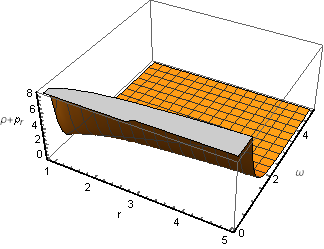}}\hspace{.1cm}
	\subfigure[The second  NEC term is found to be positive for $0.43\leq\gamma<1$, $\omega\geq 0$ and $r\geq1.4$, so, in this figure we plotted $\rho+p_t$ with respect to $r$ for $\gamma=0.5$ and $\omega\geq 0$.]{\includegraphics[scale=.65]{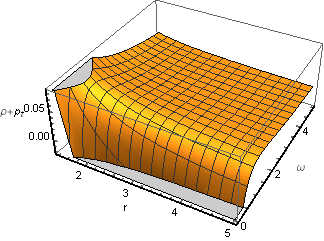}}\hspace{.1cm}
	\subfigure[The term $\rho+p_r+2p_t$ is found to be positive for $0.7\leq\gamma<1$, $\omega\geq 0$ and $r\geq1.7$. So, this figure is plotted for the term $\rho+p_r+2p_t$ with respect to $r$ for $\gamma=0.8$, $\omega\geq 0$ and $r\geq1.7$.]{\includegraphics[scale=.65]{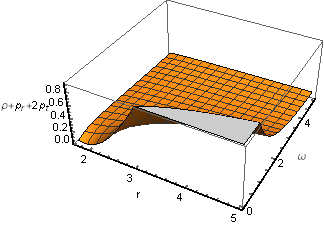}}\hspace{.1cm}
	\subfigure[This figure indicates that the behavior of the dominant energy condition term $\rho-|p_r|$ with respect to $r$ and it is found to be positive for $\gamma=0.5$.]{\includegraphics[scale=.65]{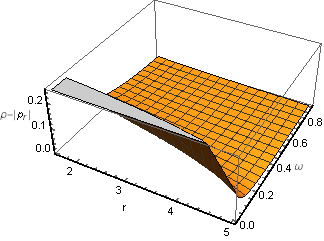}}\hspace{.1cm}
	\subfigure[This figure indicates that the behavior of the dominant energy condition term $\rho-|p_t|$ with respect to $r$ and it is found to be positive for $\gamma=0.5$.]{\includegraphics[scale=.65]{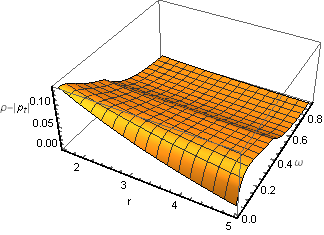}}\hspace{.1cm}\end{figure}
\begin{figure}
	\centering
		\subfigure[$\triangle$]{\includegraphics[scale=.65]{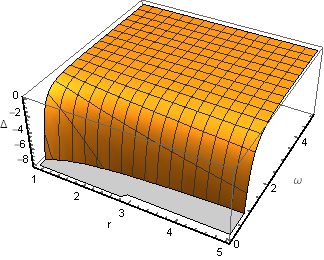}}\vspace{3cm}
	\end{figure}

\section{Results \& discussion}
In literature, violation or non-violation of energy conditions in the context of wormhole solutions is explored using the framework of various theories of  gravity. The $f(R)$ theory of gravity is one among these theories.  In this work, considering the shape function $b(r)=r_0\big(\frac{r}{r_0}\big)^\gamma$, where $0<\gamma<1$, the function $f(R)$ is derived as $f(R)=k ((\gamma+1) r (\omega+1)-(\gamma+1) \omega-1)^{\frac{\gamma^2-1}{\gamma+1}}$, where $k$ is constant and energy condition terms are computed in Section 4. For simplicity, $k=1$ is considered. This section is devoted to the analysis of energy conditions which include null energy condition (NEC), weak energy condition (WEC), strong energy condition (SEC) and dominant energy condition (DEC). The spherical regions are  determined where these energy conditions are valid. The results obtained are as follows:\\

\noindent
The energy density is examined and found to be positive for $0.43\leq\gamma<1$, $\omega\geq 0$ and $r\geq0.9$. Otherwise, it is found to be negative or imaginary. In Fig.(a), the energy density is plotted with respect to $r$ and $\omega$ taking $\gamma = 0.5$. Hence, from the figure(a), we confirm that the energy density can not be negative, if we assume the size of the throat of the wormhole is $r_0>0.9$ and the range of the parameters $\gamma$ and  $\omega$   are $0.43\leq\gamma<1$ and $\omega\geq 0$ respectively. Subsequently, the null energy condition terms are examined. The first NEC term $\rho+p_r$ is obtained to be positive for $0.43\leq\gamma<1$, $\omega\geq 0$ and $r\geq0.9$, while second  NEC term is found to be positive for $0.43\leq\gamma<1$, $\omega\geq 0$ and $r\geq1.4$. Hence, to satisfy NEC, the common region is $0.43\leq\gamma<1$, $\omega\geq 0$ and $r\geq1.4$. So, the NEC will be satisfied within this range $0.43\leq\gamma<1$, $\omega\geq 0$ and $r\geq1.4$. NEC terms are plotted in Figs.(b) \& (c)  for $\gamma = 0.5$. Subsequently, from the figure(a), (b) and (c), we observed that the WEC  will be satisfied for the range $0.43\leq\gamma<1$, $\omega\geq 0$ and $r\geq1.4$. Further, the nature of strong energy condition (SEC) term $\rho+p_r+2p_t$ is observed  and it is found to possess positive values for $0.7\leq\gamma<1$, $\omega\geq 0$ and $r\geq1.7$. This SEC term is plotted in Fig.(d) using $\gamma=0.8$, so from the figure(b), (c) and (d), it is observed that the SEC will be satisfied for the range  $0.7\leq\gamma<1$, $\omega\geq 0$ and $r\geq1.7$. Now, the dominant energy condition terms $\rho-|p_r|$ and $\rho-|p_t|$ are analyzed and found to be positive only for $0.2\leq\gamma<1$, $0\leq\omega\leq 0.9$ and $r\geq1.6$. These terms are also plotted in Figs. (e) \& (f) using $\gamma =0.5$.\\

\noindent
Hence, from the above discussion we concluded, NEC, WEC, SEC and DEC hold for $0.7\leq\gamma<1$, $0\leq\omega\leq 0.9$ and $r\geq1.7$. Validation of all energy conditions indicates that there is no exotic matter present in the region. Therefore, from this study, we could say that presence of exotic matter is not a necessary condition to construct a traversable wormhole in modified $f(R)$ gravity. That means the construction of traversable wormhole could be possible without requirement of exotic matter in modified $f(R)$ gravity. Finally, the anisotropy parameter describing the nature of geometry of wormhole is also investigated  and found to be negative for $0.2\leq\gamma<1$, $\omega\geq 0$ and $r\geq0.9$. It is also plotted in Fig. (g).  This depicts the attractive nature of geometry inside the wormhole. Thus, for $r\geq1.7$, the wormholes are filled with normal matter and possess attractive geometry. These results are also summarized in Table-1 and Table-2.

\section{Conclusion}
This work is focussed on the exploration of wormhole solutions using the framework of $f(R)$ gravity. The shape function $b(r)=r(\frac{r}{r_0})^\gamma$, where $0<\gamma<1$, defining the shape of wormhole is considered. Using this shape function and the equation of state $p_r=\omega \rho$, the function $f(R)$  is computed as $f(R)=k \bigg[(\gamma+1)^{\frac{3-\gamma}{2-\gamma}}(\omega+1)r_0^{\frac{\gamma+1}{\gamma-2}}\frac{1}{2} R^{\frac{1}{\gamma-2}}-(\gamma+1) \omega-1\bigg]^{\gamma-1}$, where $k$ is constant, which satisfies the viability conditions for $f(R)$ models. In general relativity, the violation of null energy condition (NEC) is necessary for the existence of wormhole solutions\cite{Morris95}. However, in the present study NEC is satisfied for $r\in [1.4,\infty)$. WEC, SEC and DEC all are satisfied for $r\in [1.7,\infty)$. Thus, if we consider the size of the throat $r_0$ of the wormhole  either $r_0=1.7$ or $r_0>1.7$, then all the energy conditions are satisfied throughout the wormhole geometry. Further, the wormholes possess repulsive gravitational structure filled with non-exotic matter near the throat. These observed properties of geometry, matter and energy conditions ensure the existence of wormhole solutions without violation of energy conditions away from the throat for $r\geq 1.7$ which depicts the significance of the work.
\begin{table}[!h]
	\centering
	\caption{Summary of results}
	\begin{tabular}{|c|c|l|}
		\hline
		S.No.& Terms& Results\\
		\hline
		1 & $\rho$ & $>0$, for $r\in[0.9,\infty)$, $\gamma\in[0.43,1)$, $\omega\in[0,\infty)$\\
		&        & $<0$ or imaginary, otherwise\\
		\hline
		2 & $\rho+p_r$ & $>0$, for $r\in[0.9,\infty)$, $\gamma\in[0.43,1)$, $\omega\in[0,\infty)$\\
		&        & $<0$ or imaginary, otherwise\\
		\cline{1-3}
		3 & $\rho+p_t$ & $>0$, for $r\in[1.4,\infty)$, $\gamma\in[0.43,1)$, $\omega\in[0,\infty)$\\
		&        & $<0$ or imaginary, otherwise\\
		\cline{1-3}
        4 & $\rho+p_r+2p_t$ & $>0$, for $r\in[1.7,\infty)$, $\gamma\in[0.7,1)$, $\omega\in[0,\infty)$\\
		&        & $<0$ or imaginary, otherwise\\
		\cline{1-3}
		5 & $\rho-|p_r|$ & $>0$, for $r\in[1.6,\infty)$, $\gamma\in[0.2,1)$, $\omega\in[0,0.9]$\\
		&        & $<0$ or imaginary, otherwise\\
		\cline{1-3}
        6 & $\rho-|p_t|$ & $>0$, for $r\in[1.6,\infty)$, $\gamma\in[0.2,1)$, $\omega\in[0,0.9]$\\
		&        & $<0$ or imaginary, otherwise\\
		\cline{1-3}
        7 & $\triangle$ & $<0$, for $r\in[0.9,\infty)$, $\gamma\in[0.2,1)$, $\omega\in[0,\infty)$\\
		&        & $>0$ or imaginary, otherwise\\
		\cline{1-3}
					\end{tabular}
\end{table}

\noindent
{\bf Acknowledgement:} The authors are very much thankful to the referees for their meaningful comments for the betterment of the work. The work of first author GCS is supported by CSIR Grant No.25(0260)/17/EMR-II.

\begin{table}[!h]
	\centering
	\caption{Ranges for the Satisfaction of Energy Conditions}
	\begin{tabular}{|c|c|l|}
		\hline
		S.No.& Energy Condition & Results\\
		\hline
		1 & NEC & $r\in[1.4,\infty)$, $\gamma\in[0.43,1)$, $\omega\in[0,\infty)$\\
		\hline
		2 & WEC & $r\in[1.4,\infty)$, $\gamma\in[0.43,1)$, $\omega\in[0,\infty)$\\
		\hline
		3 & SEC & $r\in[1.7,\infty)$, $\gamma\in[0.7,1)$, $\omega\in[0,\infty)$\\
		\hline
		4 & DEC & $r\in[1.6,\infty)$, $\gamma\in[0.2,1)$, $\omega\in[0,0.9]$\\
		\hline			
		5 & All ECs & $r\in[1.7,\infty)$, $\gamma\in[0.7,1)$, $\omega\in[0,0.9]$\\
		\hline
		\end{tabular}
\end{table}

\end{document}